\documentclass{aastex631}

\shorttitle{Water Evolution of Super-Earths}
\shortauthors{Moore et al.}

\graphicspath{{./}{figures/}}

\begin{document}

\title{Water Evolution \& Inventories of Super-Earths Orbiting Late M Dwarfs}

\correspondingauthor{Keavin Moore}
\email{keavin.moore@mail.mcgill.ca}

\author{Keavin Moore}
\affiliation{Department of Earth \& Planetary Sciences, McGill University, 3450 rue University, Montr\'{e}al, QC, H3A 0E8, Canada}

\author{Benjamin David}
\affiliation{Department of Physics, McGill University, 3600 rue University, Montr\'{e}al, QC, H3A 2T8, Canada}

\author{Albert Yian Zhang}
\affiliation{Department of Physics, McGill University, 3600 rue University, Montr\'{e}al, QC, H3A 2T8, Canada}

\author{Nicolas B. Cowan}
\affiliation{Department of Earth \& Planetary Sciences, McGill University, 3450 rue University, Montr\'{e}al, QC, H3A 0E8, Canada}
\affiliation{Department of Physics, McGill University, 3600 rue University, Montr\'{e}al, QC, H3A 2T8, Canada}

\begin{abstract}

%% 250-WORD LIMIT
Super-Earths orbiting M-dwarf stars may be the most common habitable planets in the Universe. However, their habitability is threatened by intense irradiation from their host stars, which drives the escape of water to space and can lead to surface desiccation. We present simulation results of a box model of water cycling between interior and atmosphere and loss to space, for terrestrial planets of mass 1--8 $M_\oplus$ orbiting in the habitable zone of a late M-dwarf. Energy-limited loss decreases with planetary mass, while diffusion-limited loss increases with mass. Depending on where it orbits in the habitable zone, a 1 $M_\oplus$ planet that starts with 3--8 Earth Oceans can end up with an Earth-like surface of oceans and exposed continents; for an 8 $M_\oplus$ super-Earth, that range is 3--12 Earth Oceans. Planets initialized with more water end up as waterworlds with no exposed continents, while planets that start with less water have desiccated surfaces by 5 Gyr. Since the mantles of terrestrial planets can hold much more water than is currently present in Earth's atmosphere, none of our simulations result in Dune planets --- such planets may be less common than previously thought. Further, more water becomes sequestered within the mantle for larger planets. A super-Earth at the inner edge of the habitable zone tends to end up as either a waterworld or with a desiccated surface; only a narrow range of initial water inventory yields an Earth-like surface.

\end{abstract}


\section{Introduction} \label{sec:intro}

\subsection{Habitability of Earth-like Planets}

Rocky planets around M-dwarfs have become prime targets for astrobiology. These stars are the most abundant in the Galaxy and are known to host many rocky planets (e.g., \citealt{dressing15, sabotta21}), but M-dwarfs are significantly more active than FGK-dwarfs \citep{scalo07, shields16}, emitting abundant X-ray and extreme ultraviolet (EUV) radiation, collectively known as XUV radiation. As a result, the habitability of rocky planets orbiting M-dwarfs is at risk due to higher irradiation during the extended early evolutionary periods of M-dwarfs (for a review, see \citealt{shields16}). 

The significant radiation impinging on the atmospheres of these rocky planets can lead to rapid and significant loss of water to space, or even surface desiccation (e.g., \citealt{wordsworth13, wordsworth14, ramirez14, luger15, tian15}). Two previous studies used box models to investigate the fate of planetary surface water as a proxy for habitability \citep{moore20, moore23}. These models include atmospheric loss to space, and track water evolution between planetary reservoirs during an early magma ocean stage and a plate-tectonics-driven deep-water cycling phase. These studies concluded that sequestering water in the solid mantle or a magma ocean can improve the habitability prospects of Earth-mass planets by reducing loss of water to space.

\subsection{Habitability of Super-Earths}

Rocky planets larger than the Earth are expected to be more common than Earth-size planets around M-dwarfs (e.g., \citealt{dressing15, mulders15, hsu20, chachan23}) and as such, their water evolution must be investigated (e.g., \citealt{kruijver21}). These ``super-Earths'' have masses in the range $1~M_\oplus < M_\mathrm{p} \lesssim 10~M_\oplus$, and super-Earths orbiting low-mass stars may be the most common habitable planets. Regardless of their intrinsic frequency, super-Earths are easier to discover and characterize than smaller terrestrial planets. Furthermore, terrestrial planets forming around low-mass stars may initially contain 10--30\% water by mass, depending on the details of their migration through the protoplanetary disk \citep{alibert17}.

How does water evolution and partitioning operate on massive terrestrial planets? The manuscript continues as follows. In Section \ref{sec:methods}, we summarize the box model of \citet{moore23} before emphasizing mass dependence and our expectations. Section \ref{sec:results} then presents temporal water evolution and parameter space results for various terrestrial planets. These results are then discussed in Section \ref{sec:discussion} along with the nuances and exclusions of our super-Earth box model, before we summarize our results and conclude in Section \ref{sec:conclusion}.

\section{Methods} \label{sec:methods}

The model used in this study is a modified version of that in \citet{moore23}, without a persistent basal magma ocean; we instead test various magma ocean water partition coefficients across different planetary masses, and a basal magma ocean would be represented by a higher partition coefficient. \citet{moore23} showed that a short-lived basal magma ocean could benefit surface habitability, which could help rapidly-desiccated planets if substantial water becomes sequestered in the mantle; however, a long-lived basal magma ocean leads to the majority of water becoming trapped in the mantle. Basal magma oceans are probably more likely for super-Earths (C.-{\'E} Boukar{\'e}, private communication, 2024), which we discuss in Section \ref{sec:discussion}. 

We direct the reader to \citet{moore23} for a thorough explanation of our model of water evolution, but we note key assumptions below. We adopt an XUV saturation timescale for the host star of $t_\mathrm{sat}=$ 1 Gyr, and use the stellar tracks of \citet{baraffe15} to determine the stellar luminosity at a given age, offsetting the track by 5 Myr to account for the lag between star and planet formation. Significant changes to the model and the mass dependence in certain equations are outlined below. The overall planetary evolution pathway remains the same as our previous study \citep{moore23}: the planet begins with a magma ocean, extending from surface to core, and which completely solidifies by the end of the runaway greenhouse phase. The runaway greenhouse phase is concurrent with the magma ocean phase. When stellar irradiation is greater than $325$ W/m$^2$ \citep{turbet21}, all water exists in a steam atmosphere; when the irradiation falls below this limit, we set the surface temperature $T_\mathrm{surf}=293.15$ K $=20^{\circ}$ C for simplicity. Following surface solidification, plate tectonics begins, initiating the deep-water cycle, i.e., exchange of water between interior and surface/atmosphere through degassing (interior to surface) and regassing (surface to interior) mediated by seafloor pressure and mantle temperature. Our deep-water cycling parameterization is based on an Earth-twin with plate tectonics \citep{moore23}.

Water loss to space is energy-limited \citep{watson81} when the planet is in a runaway greenhouse; during this time, all water is readily available in the atmosphere to be photodissociated and lost to space. Water loss to space becomes diffusion-limited \citep{walker77} once the planet exits the runaway greenhouse and the magma ocean solidifies; at this point, the more modest surface temperatures mean that water loss is limited by its vertical diffusion through the atmosphere, towards the exobase, above which it may be lost. Note that, although the diffusion limit is typically not valid for steam atmospheres, our atmosphere contains water, hydrogen, and oxygen, and hence the diffusion limit corresponds to hydrogen diffusing through an oxygen background.

\subsection{Expected Mass Dependence}

We use the scaling relations of \citet{valencia06} to calculate planetary radius, surface area, and mantle volume as a function of mass. These scalings do not explicitly account for molten phases, and thus cannot fully represent the magma ocean stage  \citep{dorn21}; however, we neglect the contribution of surface \& atmospheric water on planetary radius within our model, although we do explore this assumption in Section \ref{sec:discussion}. The relation of \citet{valencia06} between planetary radius, $R_\mathrm{p}$, and mass, $M_\mathrm{p}$ is:
\begin{equation}
    R_\mathrm{p} = R_\oplus \left(\frac{M_\mathrm{p}}{M_\oplus} \right)^{0.27},
\end{equation}
where $R_\oplus$ and $M_\oplus$ are the radius and mass of the Earth, respectively. 

We can substitute this relation into the equations for water loss to space. The energy-limited escape rate \citep{watson81} depends on the XUV energy required to overcome the gravitational potential of the planet:  
\begin{equation}\label{eqn:EL_loss}
    \dot{M}_{\mathrm{energy-limited}} = \frac{\epsilon_{\mathrm{XUV}} \pi F_{\mathrm{XUV}} R_{\mathrm{p}} R_{\mathrm{XUV}}^2}{G M_{\mathrm{p}} K_{\mathrm{tide}}}.
\end{equation}
Here, $F_\mathrm{XUV}$ is the XUV flux at the orbital distance of the planet, and we hold the orbital distance fixed throughout each simulation for simplicity; we set $K_\mathrm{p} = 1$, meaning no tidal contribution. The XUV absorption efficiency, $\epsilon_\mathrm{XUV}$, is typically 1--10\% (e.g., \citealt{ercolano10, barnes16, lopez17}), and we hold this value fixed at $\epsilon_\mathrm{XUV} = 0.1$; we previously discussed the impact of this assumption in \citet{moore23}. Making the planetary mass dependence explicit, the energy-limited escape equation becomes:
\begin{equation}
    \dot{M}_\mathrm{energy-limited} \propto R_\mathrm{p}^2 \frac{R_\mathrm{p}}{M_\mathrm{p}} = \frac{R_\mathrm{p}^3}{M_\mathrm{p}} = M_\mathrm{p}^{-0.19}.
\end{equation}
Here, we have approximated the surface area at which the XUV energy is deposited, $R_\mathrm{XUV}^2$, as $R_\mathrm{p}^2$, while the additional $R_\mathrm{p}/M_\mathrm{p}$ corresponds to the gravitational potential; see Section \ref{sec:discussion} for a sensitivity analysis of our XUV deposition radius, which may be much greater for hot, extended atmospheres, especially during the magma ocean/runaway greenhouse phase. This equation implies that the energy-limited escape rate will decrease with increasing planetary mass.

The diffusion-limited escape rate \citep{walker77} can be determined using the following equation:
\begin{equation} \label{eqn:DL_loss}
    \dot{M}_{\mathrm{diffusion-limited}} = m_{\mathrm{H}} \pi R_{\mathrm{p}}^2 \frac{b g (m_{\mathrm{O}} - m_{\mathrm{H}})}{k_{\mathrm{B}} T_\mathrm{therm} (1 + X_{\mathrm{O}}/X_{\mathrm{H}})}.
\end{equation}
Within this equation, $m_\mathrm{H}$ and $m_\mathrm{O}$ are the atomic masses of hydrogen and oxygen, respectively, the binary diffusion coefficient between hydrogen and oxygen is $b = 4.8 \times 10^{17}(T_\mathrm{therm})^{0.75}$ cm$^{-1}$ s$^{-1}$, $g$ is the planetary surface gravity, and $X_\mathrm{O}/X_\mathrm{H} = 1/2$ arises since photodissociation of water releases two hydrogen atoms for every one oxygen atom. We further assume that oxygen either escapes with hydrogen, or reacts away with the surface, such that it does not build up in the atmosphere. We adopt an average thermospheric temperature of $T_\mathrm{therm}=400$ K \citep{hunten87}, following \citet{luger15} and \citet{moore23}. We discuss the sensitivity of our results to a constant $T_\mathrm{therm}$ in Section \ref{sec:discussion}.

Again using the \citet{valencia06} scaling relations, the mass dependence of diffusion-limited escape can be clarified:
\begin{equation}
    \dot{M}_\mathrm{diffusion-limited} \propto R_\mathrm{p}^2 g \propto  R_\mathrm{p}^2 \left(\frac{M_\mathrm{p}}{R_\mathrm{p}^2}\right) =  M_\mathrm{p},
\end{equation}
so diffusion-limited escape will increase with planetary mass. The first factor of $R_\mathrm{p}^2$ again corresponds to planetary surface area. We calculate the expected water loss over 5 Gyr for each planetary mass in Table \ref{tab:diffmassmaxwaterloss}.

\begin{table*}
    \centering
    \begin{tabular}{|c|c|c|c|c|c|c|c|c}
    \hline
        Planetary & Location Within & Orbital & Orbital & \multicolumn{4}{c|}{Expected Water Loss [Earth Oceans]} \\
        \cline{5-8}
        Mass [$M_\oplus$] & Habitable Zone (HZ) & Distance [AU] & Period [days] & EL during RG & DL rest of sim & Total & Maximum \\
    \hline
         1 & Hot, Inner Edge & 0.025 & 4.81 & 7.72 & 2.82 & 10.5 & 19.7 \\
         & Middle (Mid) & 0.037 & 8.67 & 2.18 & 2.95 & 5.13 & 8.92 \\
         & Cold, Outer Edge & 0.050 & 13.6 & 0.83 & 2.99 & 3.83 & 5.66 \\
    \hline
        2 & Hot, Inner Edge & 0.025 & 4.81 & 6.77 & 5.64 & 12.4 & 17.8 \\
         & Middle (Mid) & 0.037 & 8.67 & 1.91 & 5.91 & 7.82 & 9.71 \\
         & Cold, Outer Edge & 0.050 & 13.6 & 0.73 & 5.98 & 6.71 & 7.42 \\
    \hline
        4 & Hot, Inner Edge & 0.025 & 4.81 & 5.93 & 11.3 & 17.2 & 19.4 \\
         & Middle (Mid) & 0.037 & 8.67 & 1.67 & 11.8 & 13.5 & 13.9 \\
         & Cold, Outer Edge & 0.050 & 13.6 & 0.64 & 12.0 & 12.6 & 12.8 \\
    \hline
        8 & Hot, Inner Edge & 0.025 & 4.81 & 5.20 & 22.5 & 27.8 & 27.8 \\
         & Middle (Mid) & 0.037 & 8.67 & 1.47 & 23.6 & 25.1 & 25.1 \\
         & Cold, Outer Edge & 0.050 & 13.6 & 0.56 & 23.9 & 24.5 & 24.5 \\
    \hline
    \end{tabular}
    \caption{Expected amount of water that will be lost as a function of planetary mass, $M_\mathrm{p}$, and orbital distance, neglecting a magma ocean and deep-water cycle. Each HZ orbital distance fixed throughout a given simulation, and determined using the calculator of \citet{kopparapu13} at 4.5 Gyr, comparable to the age of the Solar System; the Inner Edge corresponds to the runaway greenhouse limit, the Outer Edge corresponds to freeze-out of carbon dioxide, and the Middle is taken to be the average orbital distance between these two limits. Here, ``EL'' refers to energy-limited loss, which is expected to decrease with $M_\mathrm{p}$; ``RG'' is the runaway greenhouse; ``DL'' is diffusion-limited loss, expected to increase with $M_\mathrm{p}$; and ``Rest of Sim'' refers to the remainder of the simulation, following the end of the magma ocean/runaway greenhouse period. The total expected loss is the sum of the previous two columns, while the maximum expected loss is calculated using the maximum of EL and DL loss at each timestep. In practice, these planets lose much less water due to the cycling of water between interior, surface, and atmosphere.}
    \label{tab:diffmassmaxwaterloss}
\end{table*}

\subsection{Model Inputs \& Constraints}

Table \ref{tab:param_space} shows the parameter space explored to investigate the effect of planetary mass on water retention and water partitioning for the first 5 Gyr of the lifetime of super-Earths orbiting a late M-dwarf; we choose 5 Gyr as it is roughly the age of the Solar System, allowing direct comparison with the habitable Earth, but also to capture the early, intensely-irradiated lifetime of the planet when loss to space would be energy-limited and presumably most substantial, if not for a magma ocean, since the pre-main sequence of an M-dwarf may persist for Gyrs. Four planetary masses are investigated: $M_\mathrm{p} = 1~M_\oplus$, $2~M_\oplus$, $4~M_\oplus$, and $8~M_\oplus$. Four initial water mass fractions are explored, ranging from $4.7 \times 10^{-4}$ to $1.9 \times 10^{-3}$; these represent the end-member scenario that water delivery is a consequence of accretion, and hence, the initial water inventory will scale approximately with planetary mass. Terrestrial planets around low-mass stars may form with more than 10\% water by mass \citep{alibert17}, but we find that for water mass fractions above $2 \times 10^{-3}$, all our simulated planets end as waterworlds. 

We also run a set of simulations that begin with the same initial water inventory independent of mass, 2, 4, 6, and 8 Earth Oceans; since planets forming in-situ in the hot, volatile-depleted habitable zone around M-dwarf stars form dry (e.g., \citealt{schneeberger24}), planetary water inventory is instead delivered later by e.g., comets, and hence planets will begin with roughly the same water inventory regardless of planetary mass. Three fixed orbital distances within the habitable zone (HZ) are simulated --- Inner HZ, Mid HZ, and Outer HZ --- calculated using the HZ relations of \citet{kopparapu13} for an M8 host star at 4.5 Gyr, i.e., approximately the age of the Earth. The water saturation limit of the magma ocean, which controls the timing of atmospheric degassing, is held constant at $C_\mathrm{sat}=0.01$ due to its thorough exploration in \citet{moore23}; this means that an atmosphere is only degassed once the magma ocean becomes supersaturated with water.

\begin{table*}
    \centering
    \begin{tabular}{|c|c|c|}
    \hline
        Name & Parameter & Values Tested  \\
    \hline
         Planetary mass & $M_\mathrm{p}$ [$M_\oplus$] & 1, 2, 4, 8 \\
         Initial water mass fraction & $(M_\mathrm{init}/M_\mathrm{p})$ & $4.7 \times 10^{-4}$, $9.4 \times 10^{-4}$, $1.4 \times 10^{-3}$, $1.9 \times 10^{-3}$ \\
         Initial water inventory & $M_\mathrm{init}$ [Earth Oceans] & 2, 4, 6, 8 \\
         Magma ocean water partition coefficient & $D$ & 0.001, 0.01, 0.1 \\
         Orbital distance/location within HZ & $a_\mathrm{orb}$ [AU] & 0.025 (Inner HZ), 0.037 (Mid HZ), 0.050 (Outer HZ) \\
         Host stellar type & --- & M8 \\
    \hline
    \end{tabular}
    \caption{Parameter space explored in this study. Planetary mass is expressed in units of Earth masses, where $M_\oplus \approx 5.97 \times 10^{24}$ kg. Initial water mass fraction is based on [2, 4, 6, 8] Earth Oceans on a 1 $M_\oplus$ planet, where 1 Earth Ocean $\approx 1.4 \times 10^{21}$ kg. We run two sets of simulations: in the first, planets begin with the same initial water mass fraction, while the second initializes planets with the same mass of water, measured in Earth Oceans. These roughly represent end-members of possible water delivery scenarios.}
    \label{tab:param_space}
\end{table*}

\begin{figure}[htp]
\centering
\includegraphics[width=0.5\textwidth]{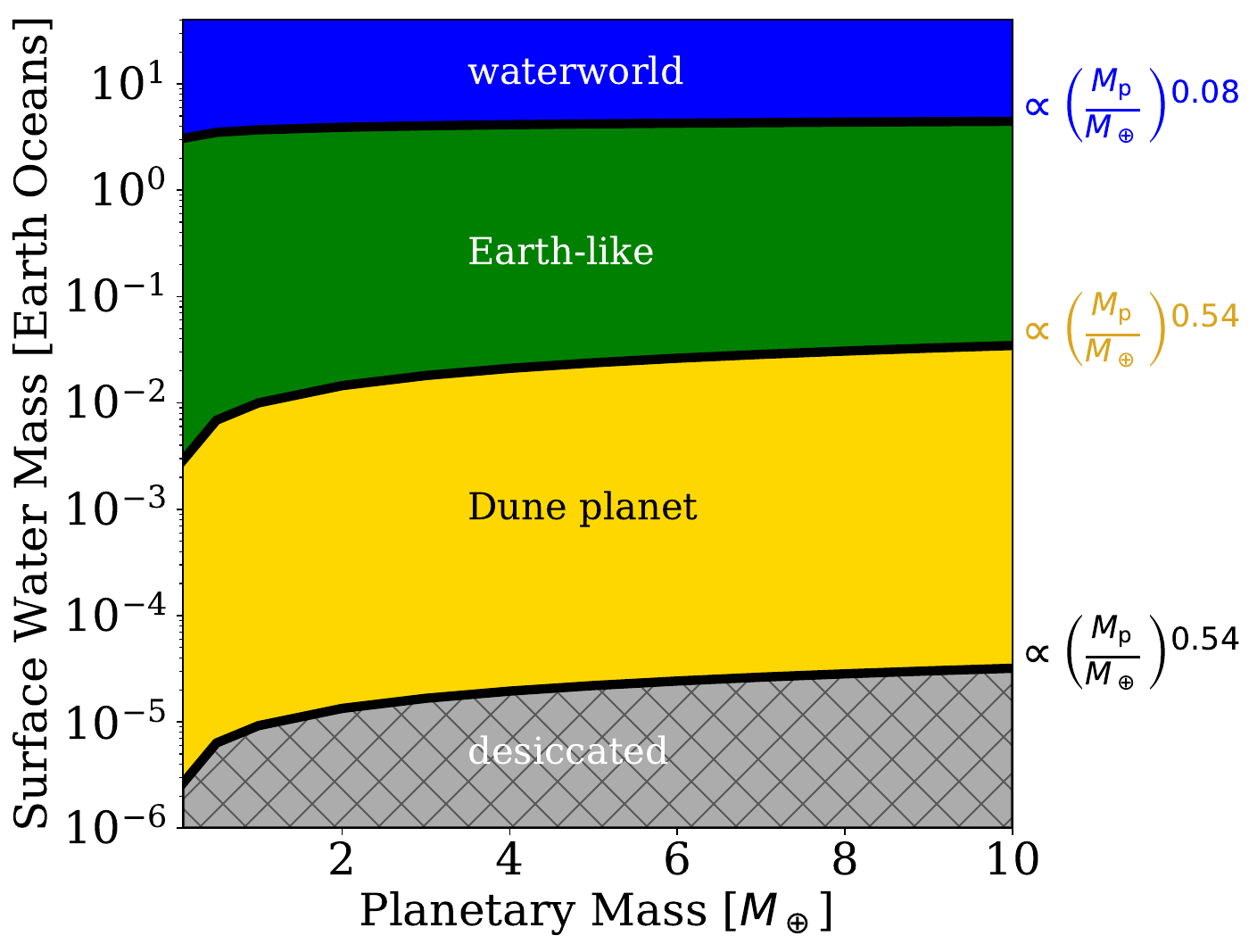}
\caption{Surface water regimes for terrestrial planets ranging from $0.1~M_\oplus$ to $10~M_\oplus$. Each limit between regimes is explained in the text. Planets in either the waterworld, Earth-like, or Dune planet regime may be habitable, while desiccated surfaces are uninhabitable. The transition between surface water regimes depends weakly on planetary mass (cf.~\citealt{cowan14}).
\label{fig:surfacewaterregimes}}
\end{figure}

The mantle water capacity will increase as planetary mass increases. As such, we adopt the 12 Earth Ocean limit on mantle water capacity for a 1 $M_\oplus$ planet from \citet{moore20} and assume the mantle water capacity scales as $(M_\mathrm{p}/M_\oplus) \times 12$ Earth Oceans, consistent with \citet{cowan14}; this is a conservative assumption given the high water capacity of post-perovskite \citep{townsend16}. Further, the different surface water regimes --- waterworld, Earth-like, Dune planet, and desiccated --- will also vary with planetary mass, as shown in Fig.~\ref{fig:surfacewaterregimes}. A waterworld will have a completely inundated surface with no exposed continents, while an Earth-like planet has both exposed continents \& surface oceans. A Dune planet would have up to ${\sim}$1\% of an Earth Ocean on its surface, while a desiccated planet has been stripped of all of its surface water, either through loss to space or regassing to the interior. These regimes can be delineated by the mass of water on the planet's surface, $W_\mathrm{s}$. A planet is a waterworld if $W_\mathrm{s} \geq 3.7 (M_\mathrm{p}/M_\oplus)^{0.08}$ Earth Oceans \citep{cowan14}, Earth-like if $0.01 (M_\mathrm{p}/M_\oplus)^{0.54} < W_\mathrm{s} < 3.7 (M_\mathrm{p}/M_\oplus)^{0.08} $ Earth Oceans, a Dune planet if $10^{-5}(M_\mathrm{p}/M_\oplus)^{0.54} \leq W_\mathrm{s} \leq 0.01 (M_\mathrm{p}/M_\oplus)^{0.54}$ \citep{abe11}, and desiccated if $W_\mathrm{s} < 10^{-5}(M_\mathrm{p}/M_\oplus)^{0.54}$ Earth Oceans, where $10^{-5}$ Earth Oceans is the amount of water currently in the Earth's atmosphere \citep{gleick93}. An Earth-like planet would maintain habitable surface conditions through the silicate weathering thermostat which regulates the climate over geological timescales (e.g., \citealt{walker81, sleep01}). A Dune planet or a waterworld, while lacking a silicate weathering thermostat, may also be habitable, which we discuss further in Sections \ref{sec:discussion} and \ref{sec:conclusion}. 

\section{Results} \label{sec:results}

\subsection{Same Initial Water Mass Fraction}

\begin{figure}[htp]
\centering
\includegraphics[width=0.47\textwidth]{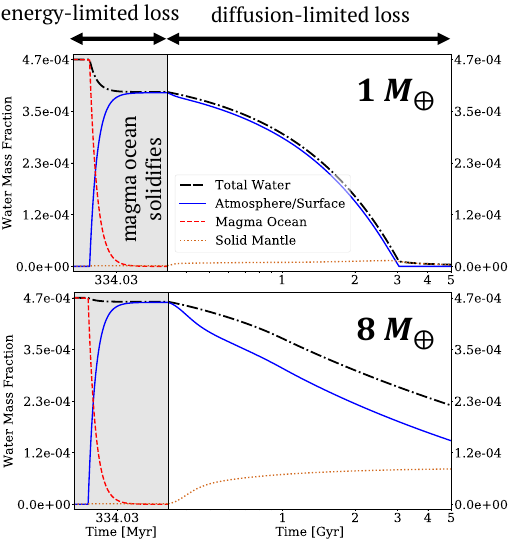}
\caption{Comparison of evolution of water inventories for terrestrial planets of mass $1~M_\oplus$ and $8~M_\oplus$. Each simulation corresponds to a planet orbiting at the inner edge of the habitable zone (Inner HZ) initialized with a water mass fraction of $4.7 \times 10^{-4}$, and assumes a magma ocean water partition coefficient of $D=0.001$. Although the $1~M_\oplus$ planet becomes desiccated, the $8~M_\oplus$ planet retains significant surface water by 5 Gyr. Although the amount of water lost to space increases with planetary mass (see Table~\ref{tab:diffmassmaxwaterloss}), so too does the initial water endowment and hence the final surface water inventory. More massive planets have a higher surface gravity, so more water becomes sequestered (and trapped) within the mantle for larger super-Earths.
\label{fig:mass_comp}}
\end{figure}

We begin with simulations initiated with the same water mass fraction (see Table \ref{tab:param_space}), which is relevant if water is delivered as part of the planetary building blocks: a $1~M_\oplus$ planet begins with [2, 4, 6, 8] Earth Oceans, while an $8~M_\oplus$ planet begins with [16, 32, 48, 64] Earth Oceans. Two such simulations are compared in Fig.~\ref{fig:mass_comp} for the same initial water mass fraction, for planets orbiting at the Inner HZ; the magma ocean/runaway greenhouse phase is shaded grey (left), while the longer deep-water cycling period is shown on the right. In Fig.~\ref{fig:mass_comp}, although more water becomes sequestered in the mantle for the super-Earth, it is a waterworld by 5 Gyr, while the $1~M_\oplus$ planet is desiccated and uninhabitable. 

We also compare the impact of magma ocean-solid mantle water partition coefficient $D$ on the water cycling between planets. As expected, $D$ controls water partitioning during the magma ocean only, by definition, and does not change the outcome after 5 Gyr. We discuss this further in Section \ref{sec:discussion}.

The results of the parameter space exploration indicate that there may be significant water present --- and potentially trapped --- within the mantle below a desiccated surface. Predictably, larger planets initiated with more water (i.e., higher water mass fraction) lead to more water trapped in the mantle and inundated, waterworld surfaces.

Overall, we find that at fixed water mass fraction, although more water is lost to space for larger super-Earths, fewer planets become desiccated and more water becomes sequestered in the mantle for larger planets. Super-Earths are more likely to become waterworlds due to their comparatively large initial water endowment with respect to the surface water regimes shown in Fig.~\ref{fig:surfacewaterregimes}.

\subsection{Same Initial Water Inventory}

We now show results for simulations that begin with the same initial water inventory regardless of mass --- [2, 4, 6, 8] Earth Oceans --- representing water delivery after planetary formation is complete, possibly through comet delivery.

\begin{figure}[htp]
\centering
\includegraphics[width=0.47\textwidth]{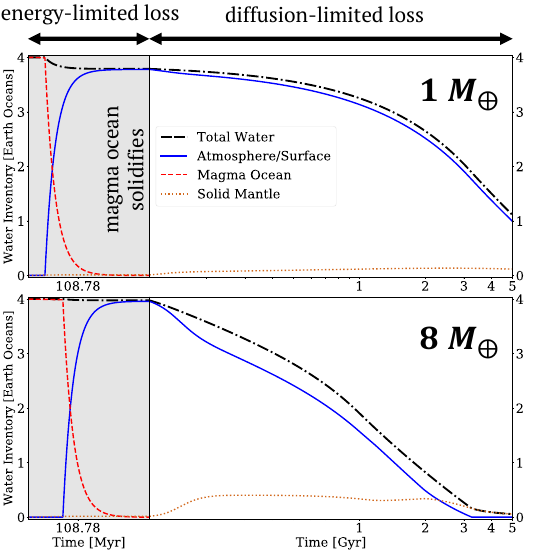}
\caption{Same as Fig.~\ref{fig:mass_comp}, but for planets initiated with the same mass of water (4 Earth Oceans) and orbiting in the middle of the habitable zone (Mid HZ). Although the $1~M_\oplus$ planet (top) ends in an Earth-like surface water regime, the $8~M_\oplus$ planet (bottom) has a desiccated, uninhabitable surface by 5 Gyr. Energy-limited loss during magma ocean decreases with planetary mass while diffusion-limited loss during deep-water cycling increases. The effect of surface gravity on mantle water inventory is visible near the beginning of the deep-water cycling period: more water becomes sequestered in the mantle of the super-Earth. Due to enhanced diffusion-limited loss, the surface of the $8~M_\oplus$ planet becomes desiccated although some water remains trapped in the solid mantle.
\label{fig:mass_comp_sameinventory}}
\end{figure}

Fig.~\ref{fig:mass_comp_sameinventory} is the same as Fig.~\ref{fig:mass_comp}, but now each planet begins the simulation with the same water inventory of 4 Earth Oceans and orbits in the middle of the habitable zone (Mid HZ). The $8~M_\oplus$ planet becomes desiccated shortly after 3 Gyr; although energy-limited loss during magma ocean is lower and larger planetary mantles sequester more water early in the deep-water cycling phase, this water is degassed to the surface and rapidly lost to space. Conversely, the $1~M_\oplus$ planet ends up in a habitable, Earth-like surface water regime.

\begin{figure*}[htp]
\centering
\includegraphics[width=0.8\textwidth]{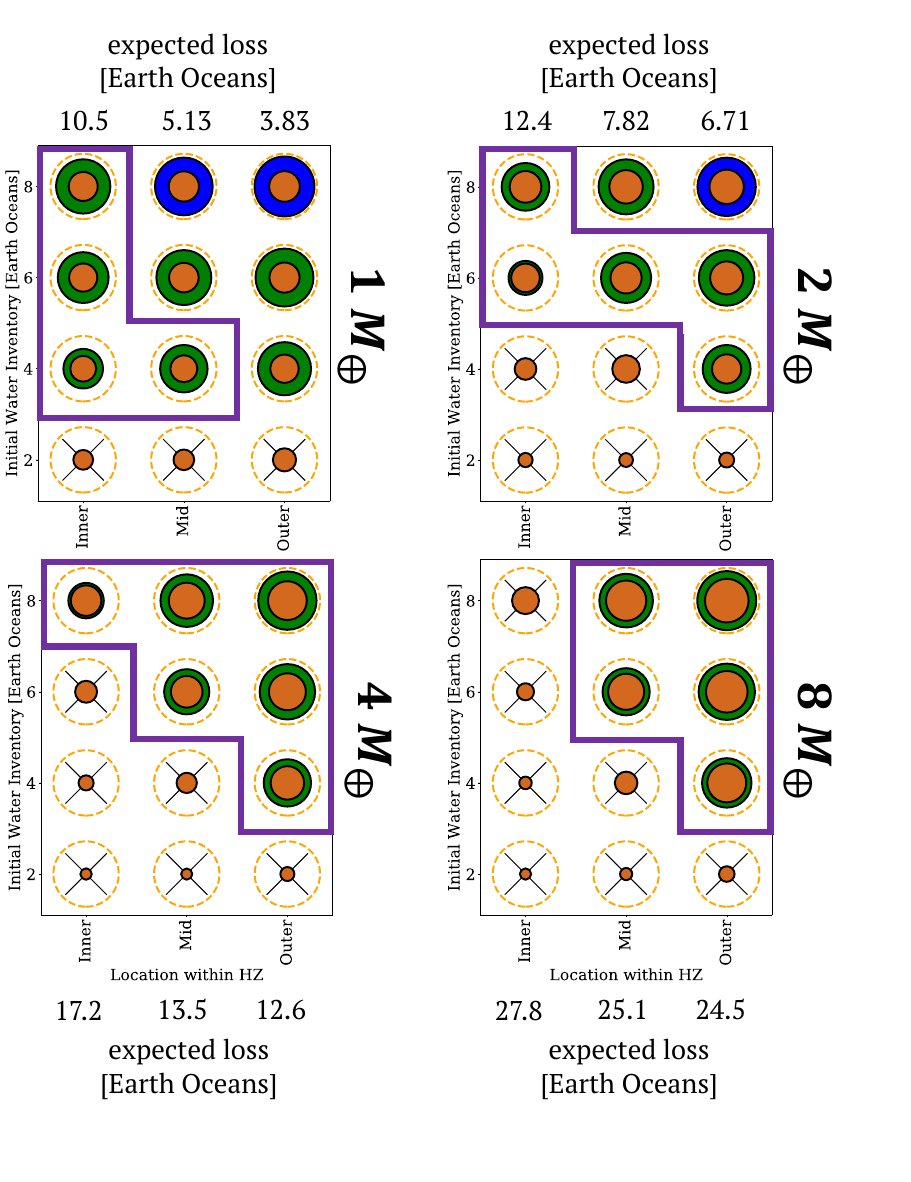}
\caption{Final water inventories for planets initiated with the same water inventory (i.e., initial water inventory independent of
mass, reasonable for the dry formation environment around M-dwarfs; see, e.g., \citealt{schneeberger24}). The initial water inventory is indicated by a dashed orange circle, and the final water inventory by a thick black circle, coloured based on the surface water regimes of Fig.~\ref{fig:surfacewaterregimes}. Additionally, the final mantle water inventory is plotted --- again scaled to the initial inventory --- as a filled brown circle. The purple boxes highlight planets that were expected to lose all of their water yet remain habitable (see text for details). Desiccated surfaces are more common on super-Earths by 5 Gyr, with fewer habitable outcomes for larger planets when initiated with fixed water mass.
\label{fig:param_search_masses_mantle_sameinventory}}
\end{figure*}

The parameter space results for fixed initial water inventory are plotted as Initial Water Inventory vs.~orbital distance in Fig.~\ref{fig:param_search_masses_mantle_sameinventory}. This figure also includes purple boxes around surface water ``survivors''. These represent planets that began the simulations with less water than they would naively be expected to lose, and so would become desiccated without a deep-water cycle (see Table \ref{tab:diffmassmaxwaterloss}), but which remain habitable in our simulations. For larger planets, more water is lost to space, and more simulations result in a desiccated surface, as indicated by a black `X'. However, over half of the simulated planets in Fig.~\ref{fig:param_search_masses_mantle_sameinventory} are in a habitable, Earth-like surface water regime by 5 Gyr. Water loss is more substantial for super-Earths due to the increase in diffusion-limited loss. Furthermore, larger planets sequester more water in their mantles due to their enhanced gravity. If water inventory is independent of planetary mass, then higher-mass terrestrial planets are more likely to be desiccated. Indeed, by 5 Gyr, there are many more desiccated planets and barely any waterworlds compared to simulations initiated with fixed water mass fraction.

\begin{figure*}[htp]
\centering
\includegraphics[width=0.9\textwidth]{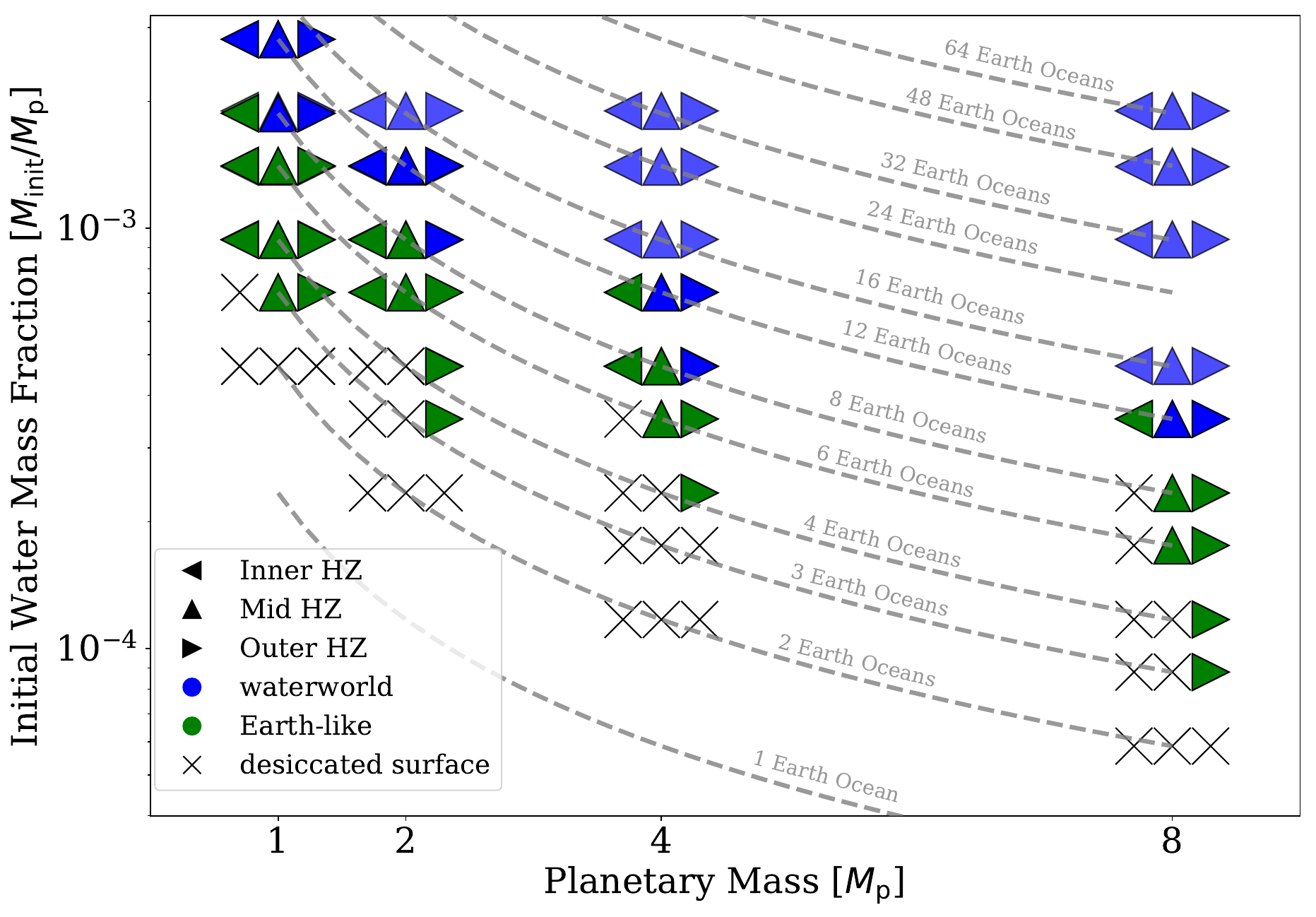}
\caption{Final surface water inventory as a function of initial water mass fraction, planetary mass, and habitable zone orbital distances (slightly offset for visualization). To first-order, the outcome of a simulation depends on the planet's initial water mass, denoted by dashed grey lines; this result is consistent with Fig.~\ref{fig:surfacewaterregimes}. To second-order, we see that at the inner edge of the HZ, a super-Earth is less habitable than an Earth-sized planet: the larger planet tends to absorb water in its mantle, in addition to losing it to space, usually resulting in a desiccated surface. 
\label{fig:WMFvsmass}}
\end{figure*}

We plot the results of all simulations for both formation scenarios as initial water mass fraction vs.~planetary mass in Fig.~\ref{fig:WMFvsmass}; here, different symbols represent the three HZ orbital positions, coloured based on their final surface water regime, and desiccated planets are represented by a black `X'. Note that we run additional simulations to fill some gaps in this figure: initial water masses of [3, 12] Earth Oceans for each planetary mass. We discuss the implications of Fig.~\ref{fig:WMFvsmass} in the following section.

\section{Discussion} \label{sec:discussion}

We have extended the $1~M_\oplus$ models of \citet{moore23} to investigate water evolution on super-Earths up to $8~M_\oplus$. We briefly describe some additional mass-dependent parameters in our model below. Many aspects, although important, are not feasible for implementation within a box model, and as such, should be explored in future studies.

Energy-limited loss was expected to decrease with increasing planetary mass, and diffusion-limited loss was expected to increase with increasing mass. These behaviours are especially apparent in Fig.~\ref{fig:mass_comp_sameinventory}, when initiating planets with the same water endowment regardless of their mass. Because the duration of diffusion-limited loss is much longer than that of energy-limited loss, the amount of water lost should increase with increasing $M_\mathrm{p}$, which we see in Fig.~\ref{fig:param_search_masses_mantle_sameinventory}; for our tested planetary masses, the decrease in energy-limited loss with mass is also consistent with the atmospheric escape calculations of \citet{chin24}. While it has been previously predicted that super-Earth atmospheres are likely stable against XUV-driven thermal escape \citep{tian09}, subsequent studies predicted immense water loss and substantial atmospheric oxygen build-up for super-Earths \citep{luger15}; however, in all three studies, planetary interior reservoirs were neglected. Furthermore, diffusion-limited loss depends on the thermospheric temperature $T_\mathrm{therm}$. We adopt $T_\mathrm{therm}=400$ K based on previous studies (e.g., \citealt{luger15}), but we note the linear dependence on $T_\mathrm{therm}$ in Eqn.~\ref{eqn:DL_loss}, such that increasing $T_\mathrm{therm}$ to ${\sim}1000$ K would result in a decrease of diffusion-limited loss by a factor of 2.5.

Our energy-limited loss parameterization presumes that the XUV deposition radius $R_\mathrm{XUV}$ is simply the planetary radius $R_\mathrm{p}$. However, $R_\mathrm{XUV}$ could be greater than $R_\mathrm{p}$, especially for a puffed-up hot, wet atmosphere above a magma ocean. We therefore test three additional XUV deposition radii to elucidate its effect on our overall results: $R_\mathrm{XUV} =$ [$R_\mathrm{p}+500$ km, $R_\mathrm{p}+1000$ km, $R_\mathrm{p}+2000$ km]. We find that the 1 $M_\oplus$ planet initiated with 4 Earth Oceans loses more water when $R_\mathrm{XUV}$ is increased, becoming desiccated by 5 Gyr, while this planet survived the original $R_\mathrm{XUV} = R_\mathrm{p}$ simulations as a habitable planet. We see this same behaviour for the 2 $M_\oplus$ planet initiated with 6 Earth Oceans; however, these are the only two simulations that are significantly changed. Regardless, our loss rates during the magma ocean/runaway greenhouse period may still be too high, as a recent study including H$_2$O photochemistry during the stellar pre-main sequence found that water loss is suppressed \citep{kawamura24}.

Super-Earths form with greater accretional energy than their Earth-mass counterparts, and likely have deeper magma oceans \citep{stixrude20}. Further, the water capacity of the early magma ocean, within which water is protected from loss to space, should increase with increasing mass (cf.~\citealt{cowan14}). A more massive planet will cool slower since volume is a stronger function of mass than surface area (e.g., \citealt{seales21}). A slowly-solidifying super-Earth may have a longer magma ocean phase than its corresponding runaway greenhouse phase. This could prolong the magma ocean phase, and maintain a hotter solid mantle during deep-water cycling. 

We note that the higher surface gravity of super-Earths may lead to the formation of high-pressure ices, inhibiting the deep-water cycling and seafloor weathering, and hence the silicate weathering thermostat; however, even our highest tested initial water inventory of 64 Earth Oceans is far below the conservative high-pressure ice limit of ${\sim}$100 Earth Oceans \citep{alibert14, nakayama19}.

We initiate every deep-water cycling simulation with the same solid mantle temperature of $T_\mathrm{m,0}=3000$ K. This value should probably scale with planetary mass, and a hotter mantle would presumably lead to more intense early degassing during deep-water cycling and, if the mantle remains hydrated, an extended period of plate tectonics (if plate tectonics begins at all on super-Earths; see, e.g., \citealt{valencia07, valencia09, korenaga10, foley12}). We can estimate $T_\mathrm{m,0}=4880$ K for an $8~M_{\oplus}$ planet by extrapolating Table 5 of \citet{schaefer15}. Using this as the initial solid mantle temperature for deep-water cycling does lead to more early degassing in our model, which slightly offsets the amount of water loss to space for a period; however, the final water inventories and partitioning remain unchanged by 5 Gyr regardless of $T_\mathrm{m,0}$. Indeed, as noted by \citet{mcgovern89} and \citet{schaefer15}, the effect of initial mantle temperature is minimal after a few hundred Myrs.

In this study we have adopted the HZ limits of \citet{kopparapu13}, but habitable planets orbit very close to their M-dwarf hosts and are likely tidally-locked (e.g., \citealt{kasting93}). Global climate model (GCM) simulations suggest that the inner edge moves inwards to 2$\times$ the stellar flux at the runaway greenhouse limit \citep{yang13, yang14}. We therefore perform a sensitivity analysis by instead taking the inner edge to be at 2$\times$ the RG limit of \citet{kopparapu13}, i.e., $a_\mathrm{Inner HZ, tl} = \sqrt{(a_\mathrm{Inner HZ}^2)/2} = 0.017$ AU. For this closer Inner HZ limit, our model predicts that the planet is permanently in a runaway greenhouse and never becomes habitable (cf. \citealt{turbet23}). In principle, a wider habitable zone incorporating a closer Inner HZ edge could benefit the super-Earths in our results (see Fig.~\ref{fig:WMFvsmass}) if the planet is able to exit the runaway greenhouse phase while retaining a water inventory.

The magma ocean water partition coefficient $D$ controls the amount of water in the solid mantle as the magma ocean solidifies. As the magma ocean solidifies from the core-mantle boundary towards the surface, the magma ocean water partition coefficient should change since it is a function of pressure \citep{papale97, papale99}; indeed, internal pressures of super-Earths will also be higher than for Earth-mass planets. We hence test $D=0.001$, 0.01, and 0.1, since we cannot fully represent this pressure-dependence in our 0-D box model as we do not resolve the 1-D internal structure. As expected, increasing $D$ leads to more substantial water sequestered in the solid mantle during the magma ocean phase.

However, $D$ only controls partitioning during the relatively short-lived magma ocean phase; less of the total water inventory may be lost to space than expected, but water sequestered in the mantle does not necessarily enhance surface habitability. All else remaining the same, the amount of water present within the solid mantle by 5 Gyr is nearly identical regardless of $D$ due to the short lifetime of the magma ocean compared to the much longer deep-water cycling phase and the significant degassing from the hot mantle following surface solidification, so varying $D$ has nearly no effect on the final planetary water inventories, their surface water regime, nor their habitability prospects. "A basal magma ocean that persists for Gyr-timescales --- posited by \citet{moore23} for an Earth-mass planet, and probably more likely to exist in super-Earths (C.~{\'E}. Boukar{\'e}, private communication, 2024) --- would presumably cause even more water to become trapped in the solid mantle than our current simulations indicate. This reinforces our conclusion that substantial water becomes locked in the solid mantle for super-Earths, in agreement with \citet{schaefer15}.

\section{Summary \& Conclusions} \label{sec:conclusion}

The surface of a super-Earth should accommodate slightly more water before becoming inundated. For the same initial water mass fraction, more massive planets are likely to avoid surface desiccation and become waterworlds with substantial water inventories trapped in their mantle; indeed, only 13/48 such simulations end in a habitable Earth-like surface water regime, while a staggering 30/48 end up as waterworlds. This is mainly due to the very large water inventories for super-Earths and the weak dependence of surface water regime on mass shown in Fig.~\ref{fig:surfacewaterregimes}. For this formation scenario, i.e., water as a consequence of accretion, although smaller planets are more likely to become Earth-like, super-Earths are less prone to desiccation and retain very wet mantles. Hence, super-Earths are possibly more likely to be habitable if waterworlds can support the origin and evolution of life. Indeed, recent studies support the long-term habitability of waterworlds in the absence of silicate weathering through CO$_2$-driven buffering \citep{ramirez18, kite18}, which can even operate in the absence of geochemical cycling; however, these results do not apply to an M8 host star, as tested throughout the current study.

Fig.~\ref{fig:param_search_masses_mantle_sameinventory}, which instead shows results for all planets initialized with 2--8 Earth Oceans, independent of planetary mass, reveals that surface desiccation is more likely for super-Earths due to the enhanced diffusion-limited loss during deep-water cycling, but also because more water becomes sequestered in the solid mantle. In this dry, in-situ formation scenario (e.g., \citealt{schneeberger24}), super-Earths are more likely to become desiccated and less likely to end up in an Earth-like surface water regime.

Although the wide swath in the log-plotted Fig.~\ref{fig:surfacewaterregimes} makes Dune planets seem like a very likely outcome of water evolution, our coarse parameter space search does not produce any. Since loss exceeds degassing late in our simulations, Dune planets may be very rare around late M-dwarfs. Although Dune planets lack a silicate weathering thermostat, the low planetary water content and dry, unsaturated atmosphere would both reduce water loss to space and widen the predicted habitable zone \citep{abe11}; since we do not find Dune planets in our parameter search, however, we leave this detail to future model iterations. Regardless, we can still estimate rough desiccation boundaries for a given planetary mass from Fig.~\ref{fig:WMFvsmass}: 1 $M_\mathrm{\oplus}$ becomes desiccated between ${\sim}$2--3 Earth Oceans, while this boundary falls between 2--8 Earth Oceans for an 8 $M_\mathrm{\oplus}$ super-Earth depending on orbital distance, with super-Earths at the Inner HZ most likely to become desiccated. Waterworlds arise for 1 $M_\mathrm{\oplus}$ above about 8 Earth Oceans, and for 8 $M_\mathrm{\oplus}$ above about 12 Earth Oceans.

Plate tectonics requires both water dissolved in the mantle and moderate surface temperatures. Our model is restricted to the plate-tectonic mode, although it is possible that super-Earths are instead in a stagnant lid regime similar to modern-day Venus, especially if surface oceans are not formed through magma ocean degassing \citep{oneill07, miyazaki22}. Insofar as stagnant lid planets degas water more effectively than they regas it, this tectonic state could help super-Earths release water that would be otherwise become trapped in their mantle by plate tectonics. However, we leave investigation of alternate tectonic modes to future modeling efforts. 

\section*{Acknowledgments}
K.~M. and N.~B.~C. thank Sean Raymond and Charles-{\'E}douard Boukar{\'e} for useful exchanges, and the anonymous referees for insightful comments and suggestions. K.~M. acknowledges support from the Natural Sciences and Engineering Research Council of
Canada (NSERC) Postgraduate Scholarships-Doctoral (PGS D) Fellowship. N.~B.~C. acknowledges support from an NSERC Discovery Grant, a Tier 2 Canada Research Chair, and an Arthur B.\ McDonald Fellowship. The authors also thank the Trottier Space Institute (TSI) and l’Institut de recherche sur les exoplan{\`e}tes (iREx) for their financial support and dynamic intellectual environment. This research was partially funded by the Heising-Simons Foundation.

\section*{Data Availability}

The model and simulation results presented within this manuscript are available in the following Zenodo repository: 10.5281/zenodo.12548229 \citep{moore24}.

\bibliographystyle{aas}

\begin{thebibliography}{99}

\bibitem[Abe et al.(2011)]{abe11} Abe, Y., Abe-Ouchi, A., Sleep, N.~H., \& Zahnle, K.~J. 2011, Astrobiology, 11, 443

\bibitem[Alibert(2014)]{alibert14} Alibert, Y.\ 2014, \aap, 561, A41. doi:10.1051/0004-6361/201322293

\bibitem[Alibert \& Benz(2017)]{alibert17} Alibert, Y. \& Benz, W.\ 2017, \aap, 598, L5. doi:10.1051/0004-6361/201629671

\bibitem[Baraffe et al.(2015)]{baraffe15} Baraffe, I., Homeier, D., Allard, F., \& Chabrier, G. 2015, \aap, 577, A42

\bibitem[Barnes et al.(2016)]{barnes16} Barnes, R., Deitrick, R., Luger, R., Driscoll, P.~E., et al. 2016, arXiv e-prints, 1608.06919v2

\bibitem[Chachan \& Lee(2023)]{chachan23} Chachan, Y. \& Lee, E.~J.\ 2023, \apjl, 952, L20. doi:10.3847/2041-8213/ace257

\bibitem[Chin et al.(2024)]{chin24} Chin, L., Dong, C., \& Lingam, M.\ 2024, \apjl, 963, L20. doi:10.3847/2041-8213/ad27d8

\bibitem[Cowan \& Abbot(2014)]{cowan14} Cowan, N.~B., \& Abbot, D.~S. 2014, \apj, 781, 27

\bibitem[Dorn \& Lichtenberg(2021)]{dorn21} Dorn, C., \& Lichtenberg, T. 2021, \apjl, 922, L4. doi:10.3847/2041-8213/ac33af

\bibitem[Dressing \& Charbonneau(2015)]{dressing15} Dressing, C.~D., \& Charbonneau, D. 2015, \apj, 807, 45

\bibitem[Ercolano \& Clarke(2010)]{ercolano10} Ercolano, B. \& Clarke, C.~J. 2010, \mnras, 402, 2735. doi:10.1111/j.1365-2966.2009.16094.x

\bibitem[Foley et al.(2012)]{foley12} Foley, B.~J., Bercovici, D., \& Landuyt, W.\ 2012, Earth and Planetary Science Letters, 331, 281. doi:10.1016/j.epsl.2012.03.028

\bibitem[Gleick(1993)]{gleick93} Gleick, P.~H. 1993, Water in Crisis: A Guide to the World's Fresh Water Resources (New York: Oxford University Press)

\bibitem[Hsu et al.(2020)]{hsu20} Hsu, D.~C., Ford, E.~B., \& Terrien, R.\ 2020, \mnras, 498, 2249

\bibitem[Hunten et al.(1987)]{hunten87} Hunten, D.~M., Pepin, R.~O., \& Walker, J.~C.~G.\ 1987, \icarus, 69, 532. doi:10.1016/0019-1035(87)90022-4

\bibitem[Kasting et al.(1993)]{kasting93} Kasting, J.~F., Whitmire, D.~P., \& Reynolds, R.~T. 1993, \icarus, 101, 108

\bibitem[Kawamura et al.(2024)]{kawamura24} Kawamura, Y., Yoshida, T., Terada, N., et al.\ 2024, \apj, 967, 95. doi:10.3847/1538-4357/ad3e7e

\bibitem[Kite \& Ford(2018)]{kite18} Kite, E.~S. \& Ford, E.~B.\ 2018, \apj, 864, 75. doi:10.3847/1538-4357/aad6e0

\bibitem[Kopparapu et al.(2013)]{kopparapu13} Kopparapu, R.~K., Ramirez, R., Kasting, J.~F., Eymet, V., et al. 2013, \apj, 765, 131

\bibitem[Korenaga(2010)]{korenaga10} Korenaga, J. 2010, \apjl, 725, L43

\bibitem[Kruijver et al.(2021)]{kruijver21} Kruijver, A., H{\"o}ning, D., \& van Westrenen, W.\ 2021, The Planetary Science Journal, 2, 208. doi:10.3847/PSJ/ac24aa

\bibitem[Lopez(2017)]{lopez17} Lopez, E.~D.\ 2017, \mnras, 472, 245. doi:10.1093/mnras/stx1558

\bibitem[Luger \& Barnes(2015)]{luger15} Luger, R., \& Barnes, R. 2015, Astrobiology, 15, 119

\bibitem[McGovern \& Schubert(1989)]{mcgovern89} McGovern, P.~J., \& Schubert, G. 1989, Earth and Planetary Science Letters, 96, 27

\bibitem[Miyazaki \& Korenaga(2022)]{miyazaki22} Miyazaki, Y. \& Korenaga, J.\ 2022, Astrobiology, 22, 713. doi:10.1089/ast.2021.0126

\bibitem[Moore \& Cowan(2020)]{moore20} Moore, K., \& Cowan, N.~B. 2020, \mnras, 496, 3786

\bibitem[Moore, Cowan, \& Boukar{\'e}(2023)]{moore23} Moore, K., Cowan, N.~B., \& Boukar{\'e}, C.-{\'E}.\ 2023, \mnras, 526, 6235. doi:10.1093/mnras/stad3138

\bibitem[Moore(2024)]{moore24} Moore, K. 2024, Model \& data associated with ``Water Evolution \& Inventories of Super-Earths Orbiting Late M-Dwarfs'', v1.0, Zenodo, doi:10.5281/zenodo.12548229

\bibitem[Mulders et al.(2015)]{mulders15} Mulders, G.~D., Pascucci, I., \& Apai, D.\ 2015, \apj, 798, 112

\bibitem[Nakayama et al.(2019)]{nakayama19} Nakayama, A., Kodama, T., Ikoma, M., \& Abe, Y. 2019, \mnras, 488, 1580

\bibitem[O'Neill \& Lenardic(2007)]{oneill07} O'Neill, C., \& Lenardic, A. 2007, Geophysical Research Letters, 34, L19204

\bibitem[Papale(1997)]{papale97} Papale, P. 1997, Contrib. Mineral Petrol, 126, 237

\bibitem[Papale(1999)]{papale99} Papale, P. 1999, American Mineralogist, 84, 477

\bibitem[Ramirez \& Kaltenegger(2014)]{ramirez14} Ramirez, R.~M. \& Kaltenegger, L.\ 2014, \apjl, 797, L25. doi:10.1088/2041-8205/797/2/L25

\bibitem[Ramirez \& Levi(2018)]{ramirez18} Ramirez, R.~M. \& Levi, A.\ 2018, \mnras, 477, 4627. doi:10.1093/mnras/sty761

\bibitem[Sabotta et al.(2021)]{sabotta21} Sabotta, S., Schlecker, M., Chaturvedi, P., Guenther, E.~W., et al. 2021, \aap, 653, A114. doi:10.1051/0004-6361/2021409

\bibitem[Scalo et al.(2007)]{scalo07} Scalo, J., Kaltenegger, L., Segura, A.~G., Fridlund, M. Ribas, I., Kulikov, Y.~N., Grenfell, J.~L., Rauer, H., Odert, P., Leitzinger, M., Selsis, F., Khodachenko, M.~L., Eiroa, C., Kasting, J., \& Lammer, H. 2007, Astrobiology, 7, 85

\bibitem[Schaefer \& Sasselov(2015)]{schaefer15} Schaefer, L., \& Sasselov, D. 2015, \apj, 801, 40

\bibitem[Schneeberger et al.(2024)]{schneeberger24} Schneeberger, A., Mousis, O., Deleuil, M., et al.\ 2024, \aap, 682, L10. doi:10.1051/0004-6361/202348309

\bibitem[Seales \& Lenardic(2021)]{seales21} Seales, J. \& Lenardic, A.\ 2021, \icarus, 367, 114560. doi:10.1016/j.icarus.2021.114560

\bibitem[Shields et al.(2016)]{shields16} Shields, A.~L., Ballard, S., \& Johnson, J.~A.\ 2016, \physrep, 663, 1. doi:10.1016/j.physrep.2016.10.003

\bibitem[Sleep \& Zahnle(2001)]{sleep01} Sleep, N.~H., \& Zahnle, K. 2001, Journal of Geophysical Research, 106, 1373

\bibitem[Stixrude et al.(2020)]{stixrude20} Stixrude, L., Scipioni, R., \& Desjarlais, M.~P.\ 2020, Nature Communications, 11, 935. doi:10.1038/s41467-020-14773-4

\bibitem[Tian(2009)]{tian09} Tian, F.\ 2009, \apj, 703, 905. doi:10.1088/0004-637X/703/1/905

\bibitem[Tian \& Ida(2015)]{tian15} Tian, F. \& Ida, S.\ 2015, Nature Geoscience, 8, 177. doi:10.1038/ngeo2372

\bibitem[Townsend et al.(2016)]{townsend16} Townsend, J.~P., Tsuchiya, J., Bina, C.~R., et al.\ 2016, Earth and Planetary Science Letters, 454, 20. doi:10.1016/j.epsl.2016.08.009

\bibitem[Turbet et al.(2021)]{turbet21} Turbet, M., Bolmont, E., Chaverot, G., et al. 2021, \nat, 598, 276. doi:10.1038/s41586-021-03873-w

\bibitem[Turbet et al.(2023)]{turbet23} Turbet, M., Fauchez, T.~J., Leconte, J., et al.\ 2023, \aap, 679, A126. doi:10.1051/0004-6361/202347539

\bibitem[Valencia et al.(2006)]{valencia06} Valencia, D., O'Connell, R.~J., \& Sasselov, D. 2006, \icarus, 181, 545

\bibitem[Valencia et al.(2007)]{valencia07} Valencia, D., O'Connell, R.~J., \& Sasselov, D.~D.\ 2007, \apjl, 670, L45. doi:10.1086/524012

\bibitem[Valencia \& O'Connell(2009)]{valencia09} Valencia, D. \& O'Connell, R.~J.\ 2009, Earth and Planetary Science Letters, 286, 492. doi:10.1016/j.epsl.2009.07.015

\bibitem[Walker(1977)]{walker77} Walker, J.~C.~G. 1977, Evolution of the Atmosphere (New York: MacMillan)

\bibitem[Walker et al.(1981)]{walker81} Walker, J.~C.~G., Hays, P.~B., \& Kasting, J.~F. 1981, Journal of Geophysical Research, 86, 9776

\bibitem[Watson et al.(1981)]{watson81} Watson, A.~J., Donahue, T.~M., \& Walker, J.~C.~G. 1981, \icarus, 48, 150

\bibitem[Wordsworth \& Pierrehumbert(2013)]{wordsworth13} Wordsworth, R.~D., \& Pierrehumbert, R.~T. 2013, \apj, 778, 154

\bibitem[Wordsworth \& Pierrehumbert(2014)]{wordsworth14} Wordsworth, R., \& Pierrehumbert, R. 2014, \apjl, 785, L20

\bibitem[Yang et al.(2013)]{yang13} Yang, J., Cowan, N.~B., \& Abbot, D.~S.\ 2013, \apjl, 771, L45. doi:10.1088/2041-8205/771/2/L45

\bibitem[Yang et al.(2014)]{yang14} Yang, J., Bou{\'e}, G., Fabrycky, D.~C., et al.\ 2014, \apjl, 787, L2. doi:10.1088/2041-8205/787/1/L2

\end{thebibliography}

\end{document}